\documentclass[12pt,preprint]{aastex}
\usepackage{emulateapj5}

\makeatletter

\newenvironment{inlinefigure}{%
\def\@captype{figure}%
\noindent\begin{minipage}{0.999\linewidth}\begin{center}}
{\end{center}\end{minipage}\smallskip}
\makeatother

\def\asca       {{\em ASCA}\/}
\def\chandra    {{\em Chandra}\/}

\def\xmm        {XMM-{\em Newton}\/}
\def\rosat      {{\em ROSAT}\/}

\usepackage{mathptmx}

\begin{document}

\title{\emph{Chandra} view of Kes 79: a nearly isothermal SNR
with rich spatial structure}

\author{
M.\ Sun,
F.\ D.\ Seward,
R.\ K.\ Smith,
P.\ O.\ Slane
}
\smallskip

\affil{Harvard-Smithsonian Center for Astrophysics,
60 Garden St., Cambridge, MA 02138; msun@cfa.harvard.edu}

\shorttitle{\chandra\ view of Kes 79}
\shortauthors{Sun et al.}

\begin{abstract}
A 30 ks \chandra\ ACIS-I observation of Kes 79 reveals rich spatial
structures, including many filaments, three partial shells, a loop and a
``protrusion''. Most of them have corresponding radio features.
Regardless of the different results from two non-equilibrium ionization
(NEI) codes, temperatures of different parts of the remnant are all
around 0.7 keV, which is surprisingly constant for a remnant with
such rich structure. If thermal conduction is responsible for smoothing
the temperature gradient, a lower limit on the thermal conductivity
of $\sim$ 1/10 of the Spitzer value can be derived. Thus, thermal
conduction may play an important role in the evolution of at least some SNRs.
No spectral signature of the ejecta is found, which suggests the
ejecta material has been well mixed with the ambient medium.
From the morphology and the spectral properties,
we suggest the bright inner shell is a wind-driven shell (WDS)
overtaken by the blast wave (the outer shell) and estimate the
age of the remnant to be $\sim$ 6 kyr for the assumed dynamics.
Projection is also required to explain the complicated morphology
of Kes 79.
\end{abstract}


\section{Introduction}

Kes 79 (G33.6+0.1) is a Galactic SNR with appreciable X-ray and radio
structure (Frail \& Clifton 1989; Seward \& Velusamy 1995).
In the radio band, there is an incomplete outer shell from the southwest
to the west with a radius of $\sim$ 6$'$ and two prominent indentations
in the east and northeast. The radio emission from the center is
bright, composed of an ``inner ring'' and many knots (Frail \& Clifton 1989).
Strong $^{12}$CO J=1$\rightarrow$0 emission and HCO$^{+}$ J=1$\rightarrow$0
emission (Scoville et al. 1987; Green \& Dewdney 1992) were found at the
system velocity of the remnant ($\sim$ 105 km s$^{-1}$), located
east and southeast of Kes 79. Green \& Dewdney (1992) conclude
that these imply an interaction of the remnant with adjacent
molecular clouds. Frail \& Clifton (1989) measured a HI kinematic
distance of 10$\pm$2 kpc for Kes 79. Case \& Bhattacharya (1998) used
a new Galactic rotation curve to revise this distance to 7.1 kpc. 

\rosat\ observations show Kes 79 has strong central shell-like emission and
diffuse faint emission extended to the radio outer shell (Seward \&
Velusamy 1995). Most X-ray emission is from a bright diffuse inner region
where  there are also bright radio filaments. \asca\ spectra of Kes
79 show strong Mg, Si, S and Fe-L lines and the global spectrum can
be fitted well by a single NEI model
(Sun \& Wang 2000; Tsunemi \& Enoguchi 2002). Kes 79 was
listed as a possible mixed-morphology SNR by Rho \& Petre (1998)
but the unusual double-shell structure (Seward \& Velusamy
1995) and X-ray / radio bright structures around the center imply
something unique.

We obtained a \chandra\ AO-2 30 ks observation of Kes 79 in 2001.
A preceding paper (Seward et al. 2003; S03 hereafter) describes
the discovery of a compact central object (CCO) which implies that
the remnant has a massive star progenitor. This paper concerns the
thermal diffuse emission. The \chandra\ data reduction is described
in $\S$2. Spatial features detected in this observation are
discussed in $\S$3. $\S$4 describes the X-ray spectral analysis. $\S$5
is the discussion and $\S$6 is the conclusion.
Throughout this paper, we use a distance d = 7.1 kpc. Thus, 1$'$ = 2.06 pc.
Dimensions, luminosities, densities, and masses scale as d,
d$^{2}$, d$^{-1/2}$, and d$^{5/2}$, respectively.

\section{\chandra\ observations \& data reduction}

The \chandra\ observation is summarized in Table 1. The aimpoint is on the
normal position of the I3 CCD. We generally follow the CIAO science
threads for the \chandra\ data reduction
\footnote{http://cxc.harvard.edu/ciao/threads/index.html}. The data were
telemetered in Faint mode and events with \asca\ grades 1, 5 and 7 were excluded.
Known bad columns, hot pixels, and CCD node boundaries also were excluded.
After removal of background flares, the effective exposure is 29.3 ks.
We applied the CXC correction for charge transfer inefficiency (CTI).
The slow gain changes in ACIS CCDs I0-I3 were also corrected using the program
`corr\_tgain' by A. Vikhlinin
\footnote{http://cxc.harvard.edu/contrib/alexey/tgain/tgain.html}.
Without this correction, offsets between
the observed line centroid and models were observed, which degrade fits
significantly. This correction significantly improves fits but has only
a small effect on best-fit temperatures (within 5\%).
An empirical correction to the ACIS low energy quantum efficiency (QE)
degradation was performed
\footnote{http://cxc.harvard.edu/cal/Links/Acis/acis/Cal\_prods/qeDeg/index.html}.
The effective area below 1.8 keV
in the front-illuminated CCDs was corrected by a factor of 0.93
to improve the cross-calibration with the back-illuminated CCDs
\footnote{http://asc.harvard.edu/cal/Links/Acis/acis/Cal\_prods/qe/12\_01\_00/}.
The response files were produced by CALCRMF and CALCARF.

It is crucial to subtract background properly for the analysis of extended
sources. The observed region is located in the Galactic plane. Thus, we
have to properly subtract the ``Galactic Ridge'' emission (e.g.,
Warwick et al. 1985; Kaneda et al. 1997). We cannot simply subtract the
standard ACIS blank sky background
\footnote{http://cxc.harvard.edu/contrib/maxim/bg/index.html} because it
is for high Galactic latitude regions. Although
parts of the ACIS-I chips are source-free, we cannot simply use the spectrum of those
regions as background spectrum because the detector response varies across
the CCDs. Therefore, we used a ``double subtraction'' method to
subtract background in the spectral analysis. Similar methods have been
applied to clusters (e.g., Markevitch \& Vikhlinin 2001). We extracted
the spectrum from regions outside of the remnant (excluding point sources).
First, we used the high-latitude blank sky background (particle background
+ Cosmic X-ray background, without ``Galactic ridge'' emission) to do the
first subtraction. The normalization of the blank sky background was
scaled up 8\% based on the observed count rate in PHA channel 2500 - 3000 ADU.
This first subtraction enables the removal of particle background from the
spectra. The residual is the vignetted ``Galactic Ridge'' emission plus
over-subtracted CXB because of the high absorption in the direction of
Kes 79. Second, we fit the residual with an
arbitrary model (power law + some gaussians). The fit is satisfactory
with $\chi^{2}$ of about 1.2. This model was taken as the ``background residual
model''. For the spectra from each region,
we first subtract the blank sky background, and then subtract the residual
background with the best-fit model derived above. The normalization
of the model is determined by the solid angle of the region of interest.
The uncertainty of the background residual model is also included in the
uncertainties of the derived properties (e.g., temperature, absorption
and abundances).

CIAO 2.2.1, FTOOLS 5.1, XSPEC 11.2 and SPEX 2.0 were used for data analysis. The
calibration data used was CALDB 2.23 from the CXC. The uncertainties in
this paper are 90\% confidence intervals unless specified otherwise.

\section{Spatial Structures}

The \chandra\ 0.5 - 3 keV exposure-corrected image of Kes 79 is shown
in Fig. 1. With the angular resolution of \chandra, filaments, multiple
shells and loops are revealed. Most of the X-ray emission is from the central
region but there is no X-ray enhancement around the central point source.
The bright inner shell is fragmented, as are the bright filaments.
Just outside the bright inner shell, there are two faint arcs which
we name ``middle shell''. The arc of this middle shell is short. It
only exists in the west and seems to merge with the inner shell in the
south. Narrow-band images centered at the Mg, Si and S emission lines were
also extracted and do not show appreciable differences from the broadband
image.

There are many similarities between the radio and X-ray images.
A radio image of Kes 79 is shown in Fig. 2. Fig. 3 shows an adaptively
smoothed exposure-corrected X-ray image with radio contours superposed.
The faint X-ray outer shell and a northeastern ``protrusion'' are clearly
revealed. The X-ray outer shell well
matches the radio outer shell, which is the expected signature of the blast-wave.
The X-ray ``protrusion'' is also coincident
with a radio ``protrusion'', which is probably a relatively low density
region between eastern clouds.
The eastern radio limb also matches X-ray filaments. Some of
the radio bright clumps around the center (but not all) also match
X-ray bright regions. At the X-ray middle shell, there is an excess
of radio emission, which may imply
compression by shocks so that the particle energy density
and magnetic field are amplified.
However, we also notice a radio-dim region filled with X-ray
bright filaments to the east of the central point source. 

The central \chandra\ point source (S03),
is close to the geometric center of the remnant in both
X-ray and radio (Fig. 2 and 3). However, it is neither at the center
defined by the radio/X-ray outer shell, nor the center defined by X-ray
bright inner shell and middle shell (Fig. 1 and 2), which is 1$'$ - 1.5$'$ to
the northeast.

Because the south-west (SW) of the remnant is more radially ordered
than other parts, we measured the radial surface brightness profiles
in two SW sectors (SW \#1 and SW \#2, Fig. 4). Three discontinuities
(the inner, middle and outer shells) are present in SW \#2, while only
two (the inner and outer shells) are present in SW \#1. Thus, there
are three apparent shells in a small portion of the remnant
(west to southwest). The surface brightness of the outer shell in
the two sectors agrees well. We assume spherical symmetry of the
X-ray emission in these two sectors and perform the deprojection
to obtain the electron density, a technique broadly applied in
galaxy cluster studies (e.g., Fabian et al. 1981). The X-ray
gas filling factor is assumed unity. The temperature and abundances
in corresponding regions are used to derive the X-ray emissivity.
The density of the outer shell is $\sim$ 0.3 - 0.4 cm$^{-3}$, while
that of the inner shell is $\sim$ 2 cm$^{-3}$ (all for a filling
factor of unity). Density jumps with
factors of 2 - 4 occur across the discontinuities.

\section{Spectral analysis}

\subsection{The average spectral properties and luminosity}

We first want to investigate the average spectral properties of Kes 79
and compare with those derived from previous observations, e.g., \asca.
The method of background subtraction was discussed in $\S$2.
The global \chandra\ 0.8 - 6 keV spectrum of Kes 79 is shown in Fig. 5.
The emission lines from He-like Ne, Mg, Si and S are significant. The
Fe-L blend is also significant as a wide bump between the Ne and Mg lines.
We first applied collisional ionization equilibrium (CIE) models
(e.g., VEQUIL in XSPEC) to fit the spectrum, but the fit is very poor
(Table 2). We then applied two NEI codes broadly used now, VNEI (version
1.1) in XSPEC 11.2\footnote{There is a version 2.0 VNEI routine in
XSPEC 11.2, which uses APED to calculate the spectrum. It produces
quite similar results as version 1.1.} and NEI in SPEX 2.0
\footnote{http://www.sron.nl/divisions/hea/spex/version2.0/}, to model
the spectra. The same abundance table (Anders \& Grevesse 1989) was used
in the two codes. Absorption, temperature, ionization timescale, and
abundances of Ne, Mg, Si, S and Fe are free parameters in models,
while the abundances of other elements are fixed at their solar
values. The fits are much better than the CIE fit (Table 2). The
two NEI codes produce somewhat different results (Table 2) but the
SPEX code fits the observed spectrum better. The best-fit SPEX NEI
model is shown in Fig. 5. The temperature derived from the SPEX model
is 15\% higher than that from the XSPEC model. The abundances of Mg, Si
and Fe from the SPEX model are also significantly higher. The discrepancy
can not be removed by setting absorptions to be the same since
decreasing the absorption in SPEX fitting will actually increase
the best-fit temperature. The different Fe-L atomic data used in
the two codes may partially account for the discrepancy (private
communication with J. Kaastra). At this stage, this systematic
difference should be kept in mind when we try to understand
the data. This discrepancy also requires caution to explain the
results of NEI fits.

We also re-analyzed the \asca\ spectrum of Kes 79 using the current
calibration files. The B2 mode data from CCD SIS0 and SIS1 were
used. Three nearby Galactic plane observations with the same CCD
mode were used as background. Source and background spectra were
extracted from the same detector region. The spectra were also fitted
by the VEQUIL, VNEI (XSPEC) and NEI (SPEX) models. Results from SIS0
and SIS1 agree so we fit them simultaneously (Table 2).
The two NEI codes still produce different results. Our results
are consistent with those of other \asca\ analyses (Sun \& Wang 2000
applying SPEX 1.1 on SIS0 data; Tsunemi \& Enoguchi 2002 applying
XSPEC 11.1 on both GIS and SIS data). \chandra\ results are
consistent with \asca\ results when the same NEI code is applied,
except that the \asca\ abundances of Mg and Si are $\sim$ 30\% higher
than \chandra\ values.

The observed flux and the luminosity of Kes 79 are derived from the
best-fit of the \chandra\ global spectrum and the total count rate.
The two NEI models still produce different results. The derived 0.5 - 10
keV flux (absorbed) is 1.2 - 1.3 $\times$ 10$^{-11}$ ergs s$^{-1}$
cm$^{-2}$. The 0.5 - 10 keV luminosity is 2.3 - 2.8 $\times$ 10$^{36}$
ergs s$^{-1}$. These values are consistent with those derived from \asca.
The CCO only contributes $\sim$ 0.13\% of the total flux.

\subsection{The spectral properties of individual regions}

We study the spectral variation across the remnant by performing
spectral analysis on 14 regions (Fig. 6). All spectra are
similar with three strong emission lines. None can be fitted well
by CIE models. We also applied the two NEI codes on these spectra.
The fits are satisfactory with $\chi^{2}_{\nu}$ ranging from 0.6
to 1.25 (for degrees of freedom from 34 to 107).  The derived
temperature and absorption distributions are shown in Fig. 6.
The temperatures derived from SPEX fits are still systematically
higher than those from XSPEC fits, while the absorption from SPEX
fits are generally $\sim$ 10\% higher. Regions 7 and 13 have the
lowest surface brightness. If we allow both the temperature and
the abundances to vary, the uncertainties are big. Thus, for
simplicity, we fix the abundances in regions 7 and 13 to the
solar values. For other regions with higher surface brightness,
we also tried to fit their spectra with solar abundances. In 
those cases, the fits can be significantly improved if the abundances
are allowed to vary.

Regardless of the systematic discrepancy of results from the two
codes, there is little temperature variation across the remnant,
no matter the geometry and location. This is surprising since
there are rich and different gas features in the remnant, which
reflects the complexity of the environment. This may give rise to
a variation of shock velocity in different regions. Thus a nearly
isothermal SNR is not expected.

The measured ionization timescales are model dependent and have
large uncertainties, but are all around 10$^{11}$ cm$^{-3}$ s$^{-1}$.
The spectral fits also show some variation of absorption. The
absorption in the west and northwest is higher than that in the
southeast. If we fix the absorption to the average value,
the change of best-fit temperature is at most 10\%.

As shown in the global fits (Table 2), the two NEI codes produce
different Mg, Si and Fe abundances. A similar discrepancy is observed
on fits for individual regions. For the ACIS spectrum with relatively
low spectral resolution, the determination of the abundance is not
so robust as that of temperature. None of the regions show more than
90\% significantly different abundances relative to the average values,
including the bright inner shell. This rules out a shocked-ejecta
nature of the bright inner shell. The X-ray bright inner shell and
clumps are more likely clouds overrun by the shock and are now
cooling and evaporating in the hot post-shock region. The failure
to find ejecta-dominant regions implies that the SN debris is
well-mixed with other material.

\section{Discussion}

\subsection{Thermal conduction}

The effect of thermal conduction in the evolution of SNRs is
controversial. Conventional wisdom says that thermal conduction
is severely suppressed in the disjointed magnetic field, which
may be common in SNRs. However, it has been argued that this effect
may not be significant in some stages of evolution (e.g.,
Cox et al. 1999; Kawasaki et al. 2002). Chevalier (1999) also
argued that molecular clouds have a significant uniform magnetic
field component so that thermal conduction is likely to be important
in the hot interior. Cox et al. (1999) proposed a model
to explain X-ray thermal-composite SNRs, which depends crucially on
efficient thermal conduction to smooth the radial temperature
gradient and increase central X-ray brightness. However, there has
been little observational evidence for efficient thermal conduction
in SNRs. This \chandra\ observation of Kes 79 may provide such evidence.

The nearly uniform temperature distribution of Kes 79 can be used to
put a lower limit on thermal conductivity. The scale length of the
temperature gradient is $l_{T}$ = $T_{e} /\mid\nabla T_{e}\mid$.
For any two distinct regions (e.g., 1 and 3 in Fig. 6), $l_{T}$ should
be at least comparable to the separation of two regions ($\sim$ 4 pc
for regions 1 and 3). The electron free path
($\lambda_{\rm e}$) in Kes 79 is 0.2 ($T_{e}$ / 0.7 keV)$^{2}$
($n_{e}$ / 1 cm$^{-3}$)$^{-1}$ pc, which is two orders of magnitude
smaller than the size of the remnant. Thus, thermal conduction should
be classical on the remnant scale. On smaller scales,
thermal conduction may be saturated especially at the interface between
the hot gas and clouds (e.g., White \& Long 1991). Assuming the Spitzer
thermal conductivity, the conduction timescale (e.g., Sarazin 1988) is:
\begin{eqnarray}
t_{\mathrm{cond}} &\equiv& -
 \left( \frac{d \ln T_{e}}{dt} \right)^{-1}
 \approx \frac{k n_{e} l_{T}^2}{\kappa} \nonumber \\
&\simeq& 38 
   \left( \frac{n_e}{1~\mathrm{cm^{-3}}} \right)
   \left( \frac{l_{T}}{10~\mathrm{pc}}\right)^2
   \left( \frac{kT_{e}}{0.7~\mathrm{keV}} \right)^{-5/2}
   \left( \frac{\ln \Lambda}{32} \right)~\mathrm{kyr},
\end{eqnarray}

where $\kappa$ is the collisional conductivity derived by Spitzer (see
the expression for $\kappa$ in e.g., Sarazin 1988), and
ln $\Lambda$ is the Coulomb logarithm, which is $\sim$ 32 for the
typical plasma temperature and density in Kes 79.
For the typical electron density in the remnant (1 cm$^{-3}$)
and typical separation of regions in Fig. 6 ($\sim$ 2$'$ or 4.1 pc), the conduction
timescale is $\sim$ 6.4 kyr, which is about the
age of the remnant ($\S$5.4). Thus, if we attribute the smooth temperature
distribution to the effect of thermal conduction, the thermal conductivity
should be close to the Spitzer value. Even if heat flux is mainly transported
in lower density regions and the uncertainty of the age is considered, the
lower limit on the thermal conductivity
is $\sim$ 1/10 of the Spitzer value.

Both the radio and X-ray morphology of Kes 79 show that it is located in a
non-uniform environment and is probably interacting with clouds, which
should produce complicated magnetic field structure.
Thus, we need to understand why thermal conduction is not significantly
suppressed. Narayan \& Medvedev (2001) show that thermal conductivity
can approach one fifth of the Spitzer value if the magnetized plasma has
chaotic magnetic field fluctuations extended over two or more decades in
length scales. Numerical work by Chandran \& Maron (2003) also shows that
thermal conduction is only reduced by a factor of 5-10 relative to
the Spitzer value in the static field approximation. Cho et al. (2003)
used numerical methods to study thermal conduction in magnetized turbulent
gas and found no suppression. Thus, from theoretical points of view, thermal
conductivity on the level of 1/10 of the Spitzer value can be achieved
in SNRs.

Kes 79 may not be the only SNR with little temperature variation.
\asca\ observations of G69.7+1.0 (Yoshita et al. 2000), 3C400.2 (Yoshita
et al. 2001) and 3C391 (Chen \& Slane 2001) also show little
temperature variation across the remnants. However, these \asca\ observations
only yield spectral information from 2 - 4 regions (compared to 14
regions used in this work).
Moreover, none of them appear to display the rich structure seen
in Kes 79. \chandra\ and \xmm\ observations could better constrain
the temperature variations in these remnants.

\subsection{Multi-shell structure}

The \chandra\ image of Kes 79 clearly shows multi-shell structure,
as do the radio images, especially at 5 GHz (Fig. 3 of
Velusamy et al. 1991). The radio middle shell is at the same
azimuth of the X-ray middle shell, but more radially outward.
The good correspondence of radio and X-ray implies that these
shells are caused by shock compression. As shown in the early
analysis, none represents the reverse shock. It is unlikely that
the ordered shapes of inner and middle shells are
caused by engulfed evaporating cloudlets behind the blast wave.
There are at least two possibilities that may cause the multi-shell
structure.

The first is projection. Tenorio-Tagle,
Bodenheimer \& Yorke (1985) made simulations on the interaction
of SNRs with molecular clouds. If a SN explodes on the outskirts
of a molecular cloud, after the shock reaches the edge of the cloud
and breaks out, the shock will speed up in the lower density intercloud
medium, while the shock in the cloud is still moving slowly. Thus, in
the simplest case, we might see two half shells with different radii,
where the radius ratio depends on the initial location of the SN event
and the density gradient. If the line of sight is perpendicular to
the edge of the cloud, we will see the two shells overlap on the sky
creating the appearance of a double shell. The heavy element abundances of
the two shells will be determined by the ISM abundance. This can roughly
explain the outer and inner shells and their abundances.

An alternative explanation requires the presence of a wind-driven
shell (WDS). The compact source found in the center of Kes 79
(S03) implies a massive progenitor. Such massive stars are known
to have strong stellar winds during their short lifetimes. Thus,
Kes 79 may have begun to evolve inside the stellar wind bubble of
its progenitor. The presence of a WDS has a significant effect on
the evolution and appearance of an SNR. If the WDS is not very massive,
when the SN shock hits the WDS, the whole shell expands while the
fast-moving blast wave overtakes it and propagates
into the undisturbed ambient medium. At the same time,
a reflected shock is expected to form and propagate inwards.
This results in multi-shell structure of the remnant, with complicated
density, temperature and velocity distributions (Tenorio-Tagle et al.
1990, 1991). Severe distortion of the WDS by cooling and
Rayleigh-Taylor instabilities is also revealed in 2-D simulations
(Tenorio-Tagle et al. 1991). In this picture, the inner shell of
Kes 79 is interpreted as an overtaken WDS, evaporating in the interior
of the remnant. The opening of the inner shell at the north may indicate
a blow-out of the wind, or could be due to the motion of the progenitor towards
the south. The outer shell is the blast wave running in the
undisturbed medium. The blast wave should have been decelerated when
crossing the WDS. The middle shell could be the projected outer shell in
another direction where the blastwave has crossed a locally denser medium.
A local reflection shock is another possibility.

In summary, both scenarios can roughly explain the observed properties.
We prefer the latter since it is a natural consequence of massive star
evolution. Projection is still needed to explain the complicated
morphology.

However, neither scenario explains the ionization timescales of the
inner shell and the outer shell derived from the NEI fitting.
The outer shell and inner shell seem to have similar ionization
timescales, implying that the denser inner shell is more recently shocked
than the outer shell. Both models require
the outer shell to be shocked at a later or comparable time than the inner
shell. The current spectral measurement of the faint outer shell still,
however, has large uncertainties. Future observations may resolve this puzzle.

\subsection{Filaments}

At least 7 bright or faint clumpy filaments are found in this observation
(Fig. 1). They have a width of 6 - 9 arcsec (0.2 - 0.3 pc), compared to their
length of 2 - 3 arcmin (3.1 - 6.2 pc). We measured the surface brightness
of the two eastern bright filaments and their surroundings. The filaments
are brighter by a factor of 12 - 15, implying an average density
difference of $\sim$ 4. This difference is just as predicted for adiabatic
shocks. Thus, these two filaments are probably thin shells viewed edge-on.
They may also be part of the shocked WDS.
There are several filaments in the interior,
however, for which this interpretation is less possible. Filament-like
structures are also predicted by numerical simulations of shock-cloud
interactions (e.g., Xu \& Stone 1995), but in simulations, the cloud
is fragmented and filament-like structures are very complicated and
chaotic. The filaments in the NE sector of Kes 79 seem to be very ordered.
This observation provides a challenge to model the interior structure
in middle-aged SNRs.

\subsection{Age, density and mass of Kes 79}

Accurate estimate of age depends on the understanding of the dynamical
evolution of the remnant. The simple Sedov solution and the cloudlet
evaporation scenario proposed by White \& Long (1991) cannot fit the
observed properties. In $\S$5.2, we suggest that the bright inner shell may be
the overrun WDS, which complicates the dynamical history
of the remnant. As pointed out and discussed by Weaver et al. (1977),
the WDS can be considered as a thin shell after the cooling in the WDS
is important. Chen et al. (2003) derived a semi-analytic solution for
SNRs which evolve crossing a density jump in the surrounding medium,
also assuming a thin shell. For Kes 79, since the radius of inner shell
is small ($\sim$ 6 pc), the remnant should be in the free expansion stage
before the blastwave hits the WDS. After the hit, the remnant is in the
Sedov phase. Thus, we can apply the solution of Chen et al. (2003) to Kes 79.
Using a blastwave temperature (T$_{\rm S}$) of 0.6 keV, a density ratio
$\beta$ of 0.1 (see
the definition in Chen et al. 2003), radii of 3$'$ (6.2 pc) and 6$'$
(12.4 pc) for the inner and outer shells respectively, the time after the
blastwave hits the WDS is 5.2 (T$_{\rm S}$/0.6 keV)$^{-1/2}$ kyr, from
equ. 7, 8, 10 and 11 in Chen et al. (2003). This result is not sensitive to
the choice of $\beta$. If the velocity of the blastwave
is 5000 - 10000 km s$^{-1}$ inside the WDS, the age of the remnant is 5.8 -
6.4 kyr. For T$_{\rm S}$ in the range of 0.4 - 0.7 keV, the age is 5.4
- 7.5 kyr. Not surprisingly, the derived age is close to the one derived
from the Sedov solution (5.9 - 7.8 kyr for a shock temperature of 0.4 -
0.7 keV; Sedov 1959). A lower limit on the age of the remnant can also
be derived from the measured ionization timescale. The ionization
timescales derived for individual small regions have large uncertainties,
but are all around 10$^{11}$ cm$^{-3}$ s. Using an average electron
density of 0.5 - 1 cm$^{-3}$, we set a lower limit on the age of remnant:
3.2 - 6.4 kyr, which is consistent with the estimate from dynamics.

The ambient density (n$_{0}$) can also be estimated from the work of Chen et al.
(2003), n$_{0}$ = 0.36 E$_{51}$ (T$_{\rm S}$/0.6 keV)$^{-1}$ cm$^{-3}$ for the
assumed shock radius and WDS radius, where
E$_{51}$ is the explosion energy in the unit of 10$^{51}$ ergs s$^{-1}$.
This may be only the average value, since the large brightness contrast
between the inner shell and outer shell may imply a radial ambient density
gradient. If we assume a spherical and uniform distribution of the ambient
medium, this density corresponds to a swept-up ISM mass of 99 E$_{51}$
(T$_{\rm S}$/0.6 keV)$^{-1}$ M$_{\odot}$. We also can estimate the
average electron density and mass of the X-ray emitting gas from the best-fit
emission measure. Since the regions within the inner 3$'$ radius and at 3$'$ -
6$'$ have large surface brightness contrast, the gas properties are derived
separately in the two regions. The different emission measure derived from
two NEI codes (Table 2) adds some uncertainty here. The projection from the
outer bin to the inner bin is estimated by assuming spherical symmetry. The
results are: for the inner 3$'$, $\bar{n_{e}}$ = 0.8 - 1.2 f$^{-1/2}_{1}$
cm$^{-3}$ and M$_{\rm X}$ = 28 - 40 f$^{1/2}_{1}$ M$_{\odot}$, for the
region between 3$'$ and 6$'$, $\bar{n_{e}}$ = 0.25 - 0.35 f$^{-1/2}_{2}$
cm$^{-3}$ and M$_{\rm X}$ = 44 - 62 f$^{1/2}_{2}$ M$_{\odot}$, where f$_{1}$
and f$_{2}$ are the filling factors of the hot gas in the inner and outer
bins respectively. The inner bin (dominant by bright shells and filaments)
and the outer bin (mostly faint diffuse emission) should have different
filling factors. But in any cases, it is clear that all the X-ray emitting
gas can be the swept-up ISM.

\section{Conclusion}

The \chandra\ observation of SNR Kes 79 has led to the following
conclusions:

1. The \chandra\ image of Kes 79 reveals rich spatial structures:
filaments, multiple shells, a loop and a ``protrusion'', implying
a complicated environment around Kes 79. Most of the X-ray structures
have corresponding radio structures, implying the importance of
shock compression.

2. Three shells are detected in the SW part of the remnant.
The inner shell may be an evaporating WDS overrun by the blast wave (the
outer shell). The middle shell is perhaps due to projection or a reflection
shock. The age of the remnant is estimated to be $\sim$ 6 kyr from the
assumed dynamics.

3. We do not find regions with significantly enriched heavy elements,
including the bright inner shell. This implies that the ejecta have
been mixed well with the surrounding medium given the relative youth
of the remnant.

4. Regardless of the systematic discrepancy of results from the two
codes, there is little temperature variation across the remnant,
no matter the geometry and location. If thermal conduction
is responsible for smoothing the temperature gradient, a lower limit
of $\sim$ 1/10 Spitzer value on the thermal conductivity can be
obtained. Thus, thermal conduction may have a significant effect in
the evolution of SNRs.

\acknowledgments

We acknowledge inspiring discussions with Y. Chen. We thank the
anonymous referee for prompt and valuable comments.
This work was supported by NASA Grant GO1-2067X.

\clearpage

\vspace{1.5cm}
\begin{center}
TABLE 1
\vspace{3mm}

{\small
{\sc The \chandra\ observation}
\vspace{2mm}

\begin{tabular}{lc}
\hline\hline

Observation ID & 1982 \\
Observation mode & ACIS-I, 012378 \\
Data mode & Faint \\
Date & July 31 - August 1, 2001 \\
Total exposure (ks) & 29.95 \\
Effective exposure (ks) & 29.27 \\
Counting rate (0.5 - 10 keV) (cts/s) & 1.7 \\

\hline\hline
\end{tabular}}
\end{center}

\vspace{1.5cm}
\begin{center}
TABLE 2
\vspace{3mm}

{\small 
{\sc The fit to the global spectrum of Kes 79}
\vspace{2mm}

\begin{tabular}{ccccccc}
\hline \hline
 &\asca\ $^{\rm a}$ &\asca\ $^{\rm b}$ & \asca\ $^{\rm c}$& \chandra\ $^{\rm a}$ &\chandra\ $^{\rm b}$& \chandra\ $^{\rm c}$\\
 & VEQUIL & VNEI & NEI & VEQUIL & VNEI & NEI \\\hline

N$_{\rm H}$ (10$^{22}$ cm$^{-2}$) & 1.70$\pm$0.05 & 1.63$^{+0.06}_{-0.05}$ & 1.78$\pm$0.05 & 1.58$\pm$0.04 & 1.54$^{+0.05}_{-0.04}$ & 1.67$\pm$0.04 \\
T (keV) & 0.56$^{+0.02}_{-0.01}$ & 0.66$^{+0.03}_{-0.02}$ & 0.78$\pm$0.03 & 0.57$\pm$0.01 & 0.66$\pm$0.02 & 0.76$\pm$0.03 \\
$\tau$ (10$^{11}$ cm$^{-3}$ s) & - & 0.90$^{+0.30}_{-0.19}$ & 0.88$^{+0.13}_{-0.11}$ & - & 0.78$^{+0.24}_{-0.16}$ & 0.98$^{+0.15}_{-0.13}$ \\
Ne & 1.55$^{+0.80}_{-0.58}$ & 0.52$^{+0.16}_{-0.15}$ & 0.59$^{+0.32}_{-0.28}$ & 0.94$^{+0.33}_{-0.28}$ & 0.42$\pm$0.11 & 0.31$\pm$0.17 \\
Mg & 1.64$^{+0.38}_{-0.27}$ & 0.73$^{+0.07}_{-0.06}$ & 1.87$^{+0.24}_{-0.20}$ & 1.31$^{+0.18}_{-0.15}$ & 0.60$\pm$0.04 & 1.46$^{+0.14}_{-0.12}$ \\
Si & 1.50$^{+0.29}_{-0.22}$ & 0.79$^{+0.07}_{-0.06}$ & 1.33$^{+0.14}_{-0.13}$ & 1.03$^{+0.11}_{-0.10}$ & 0.57$\pm$0.04 & 1.01$^{+0.09}_{-0.08}$ \\
S & 2.04$^{+0.41}_{-0.32}$ & 1.08$^{+0.15}_{-0.14}$ & 1.25$^{+0.17}_{-0.15}$ & 1.78$^{+0.21}_{-0.20}$ & 1.04$^{+0.15}_{-0.12}$ & 1.03$^{+0.12}_{-0.11}$ \\
Fe & 1.41$^{+0.35}_{-0.25}$ & 0.59$^{+0.09}_{-0.09}$ & 1.92$^{+0.35}_{-0.31}$ & 1.09$^{+0.17}_{-0.15}$ & 0.48$\pm$0.06 & 1.52$^{+0.22}_{-0.20}$ \\
Norm$^{\rm d}$ & 5.16$\pm$0.74 & 6.29$^{+0.70}_{-0.65}$ & 2.94$^{+0.31}_{-0.29}$ & 5.02$\pm$0.42 & 5.95$^{+0.67}_{-0.49}$ & 3.02$^{+0.24}_{-0.23}$ \\
$\chi^{2}$/dof & 396.2/182 & 236.9/181 & 210.0/175 & 480.1/163 & 296.9/162 & 181.9/136 \\

\hline\hline
\end{tabular}
\begin{flushleft}
\leftskip 25pt
$^{\rm a}$ Ionization equilibrium collisional plasma model in XSPEC 11.2 \\
$^{\rm b}$ Non-equilibrium ionization collisional plasma model (version 1.1) in XSPEC 11.2 \\
$^{\rm c}$ Non-equilibrium ionization collisional plasma model in SPEX 2.0 \\
$^{\rm d}$ Normalization of the model (n$_{\rm e}$ n$_{\rm H}$ V) in units of
10$^{58}$ cm$^{-3}$. We used a distance of 7.1 kpc. Note these values are $\sim$ 30\%
less than those of the whole remnant. The big difference in the normalization
between XSPEC and SPEX results comes from their difference on modeling the line
emission and continuum.  \\
\end{flushleft}}
\end{center}

%%%%%%%%%%%%%%%%%%%%%%%%%%%%%%%%%%%%%%%%%%%%%%%%%%%%%%%%%%%%%%%%%%%%%%%%%%
\begin{inlinefigure}
\vspace{-1.7cm}
  \centerline{\includegraphics[height=1.0\linewidth]{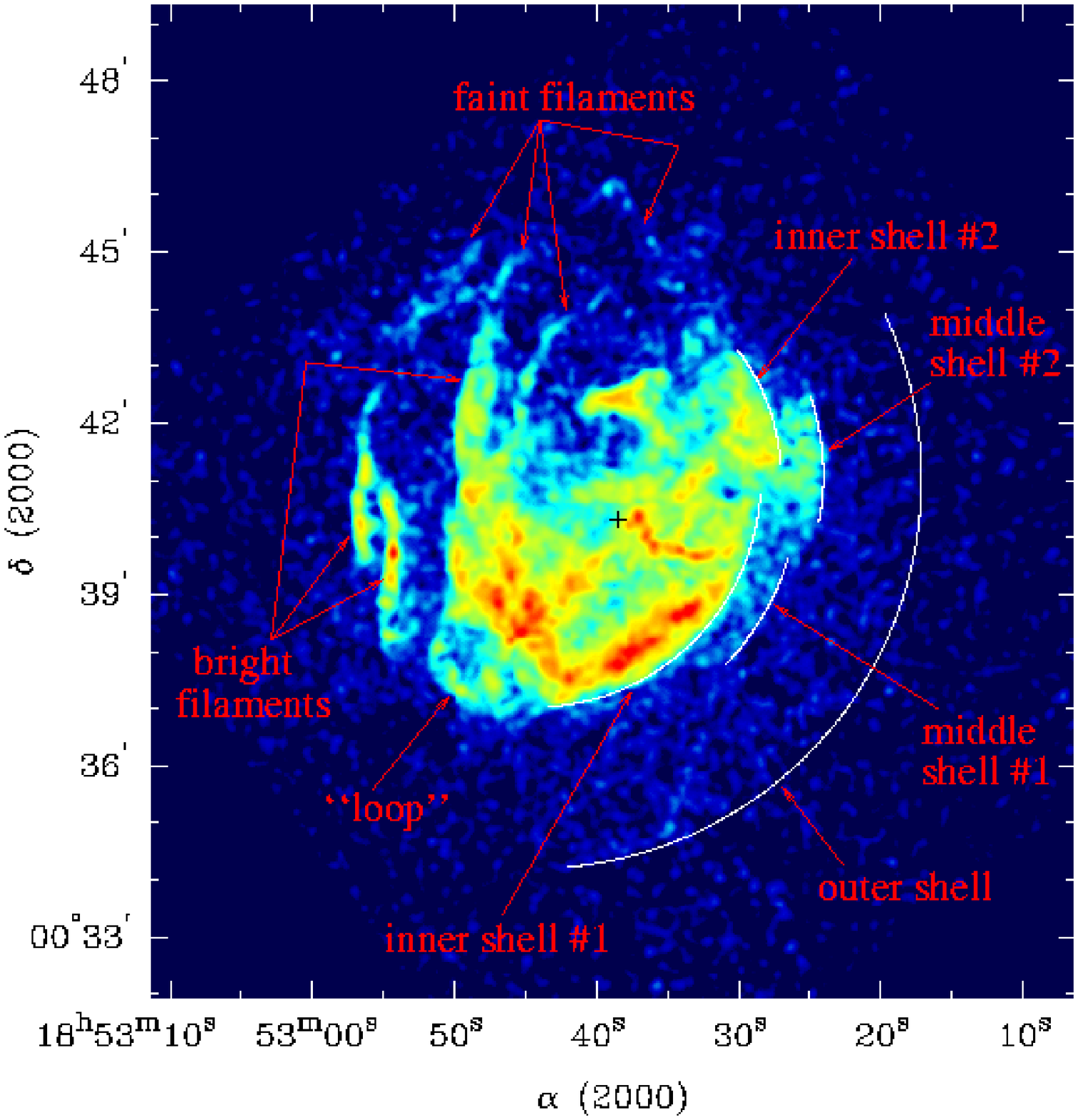}}
\vspace{-1.7cm}
  \caption{The 0.5 - 3 keV \chandra\ image of Kes 79 (exposure-corrected),
smoothed by a Gaussian with $\sigma$ of 4$''$. Point sources were removed.
Interesting features, including partial shells and filaments, are marked.
The black cross shows the position of the central point source (S03).
    \label{fig:img:smo}}
\end{inlinefigure}
%%%%%%%%%%%%%%%%%%%%%%%%%%%%%%%%%%%%%%%%%%%%%%%%%%%%%%%%%%%%%%%%%%%%%%%%%%

%%%%%%%%%%%%%%%%%%%%%%%%%%%%%%%%%%%%%%%%%%%%%%%%%%%%%%%%%%%%%%%%%%%%%%%%%%
\vspace{1cm}
\begin{inlinefigure}
  \centerline{\includegraphics[height=0.8\linewidth,angle=270]{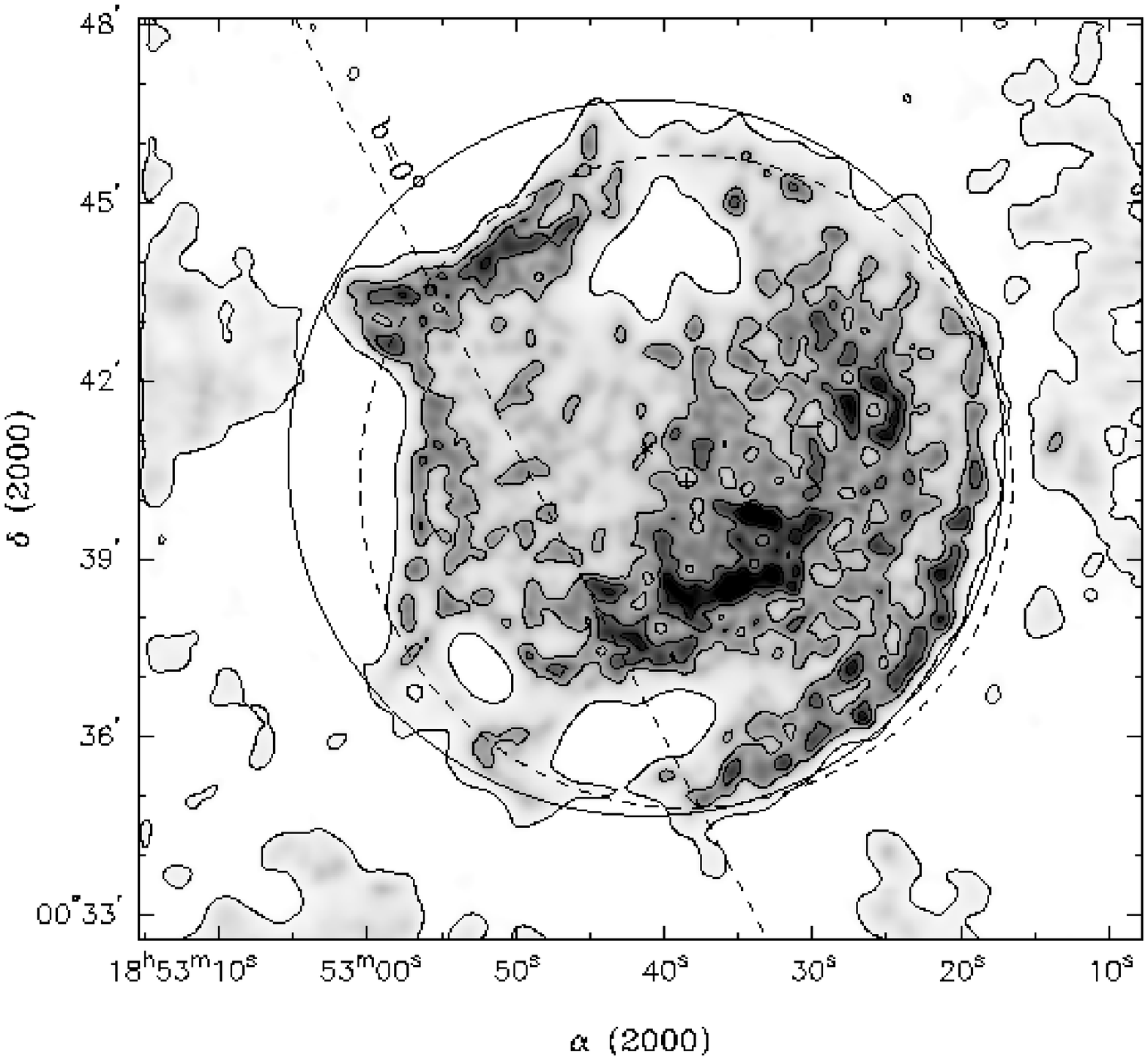}}
  \caption{VLA 20 cm map of Kes 79 (Velusamy, Becker \& Seward 1991)
superposed on its own contours. The position of the central X-ray point
source is shown by a cross. The large dashed circle is centered at the central
point source with a radius of 5.5$'$. The curvature of the radio outer
shell does not match a circle centered at the point source. The large
solid circle (with a radius of 6$'$) matches the radio boundary in the
N, E, and S well. Its center is marked
by an asterisk. The two centers are $\sim$ 1$'$ apart.
    \label{fig:img:smo}}
\end{inlinefigure}
%%%%%%%%%%%%%%%%%%%%%%%%%%%%%%%%%%%%%%%%%%%%%%%%%%%%%%%%%%%%%%%%%%%%%%%%%%

%%%%%%%%%%%%%%%%%%%%%%%%%%%%%%%%%%%%%%%%%%%%%%%%%%%%%%%%%%%%%%%%%%%%%%%%%%
\vspace{-1cm}
\begin{inlinefigure}
  \centerline{\includegraphics[height=1.0\linewidth]{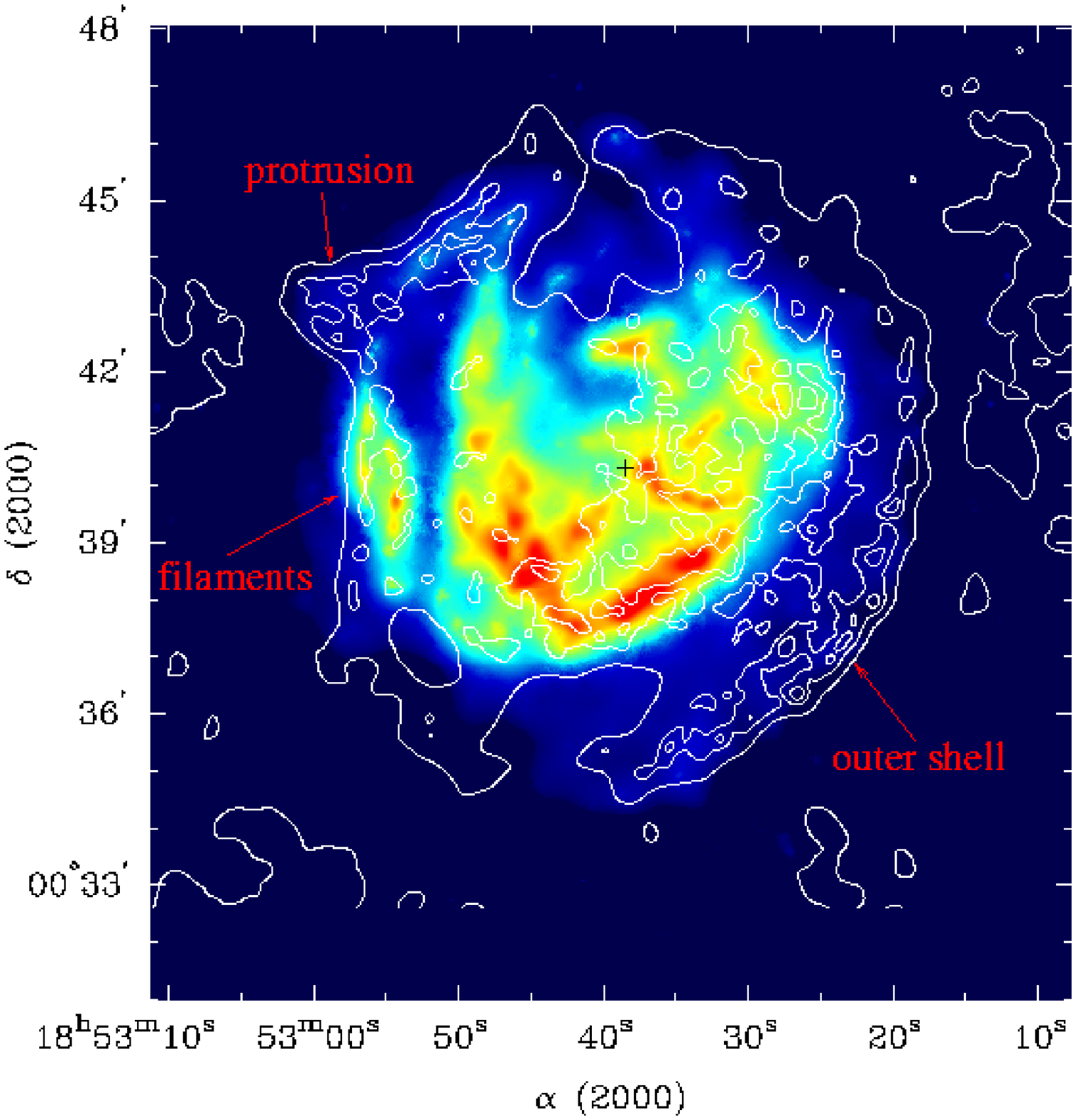}}
\vspace{-1.7cm}
  \caption{\chandra\ image in the 0.5 - 3 keV band (exposure-corrected,
adaptively smoothed, shown in square-root scale), superposed on VLA 20
cm contours. The smoothing scale ranges from 6$''$ to 20$''$ depending
on the observed photon statistics. The X-ray outer shell and ``protrusion''
are better shown in this figure than in Fig. 1.
X-ray and radio structures are generally coincident with
each other, e.g., the outer shell, part of the bright inner shell, some
filaments and the ``protrusion''. However, there is a radio-dim, X-ray bright
region in the east interior to the radio boundary which is apparently part of the
inner ring. The black cross shows the position of the central point
source.
    \label{fig:img:smo}}
\end{inlinefigure}
%%%%%%%%%%%%%%%%%%%%%%%%%%%%%%%%%%%%%%%%%%%%%%%%%%%%%%%%%%%%%%%%%%%%%%%%%%

%%%%%%%%%%%%%%%%%%%%%%%%%%%%%%%%%%%%%%%%%%%%%%%%%%%%%%%%%%%%%%%%%%%%%%%%%%
\begin{figure*}[htb]
\vspace{-4cm}
  \centerline{\includegraphics[height=1.2\linewidth]{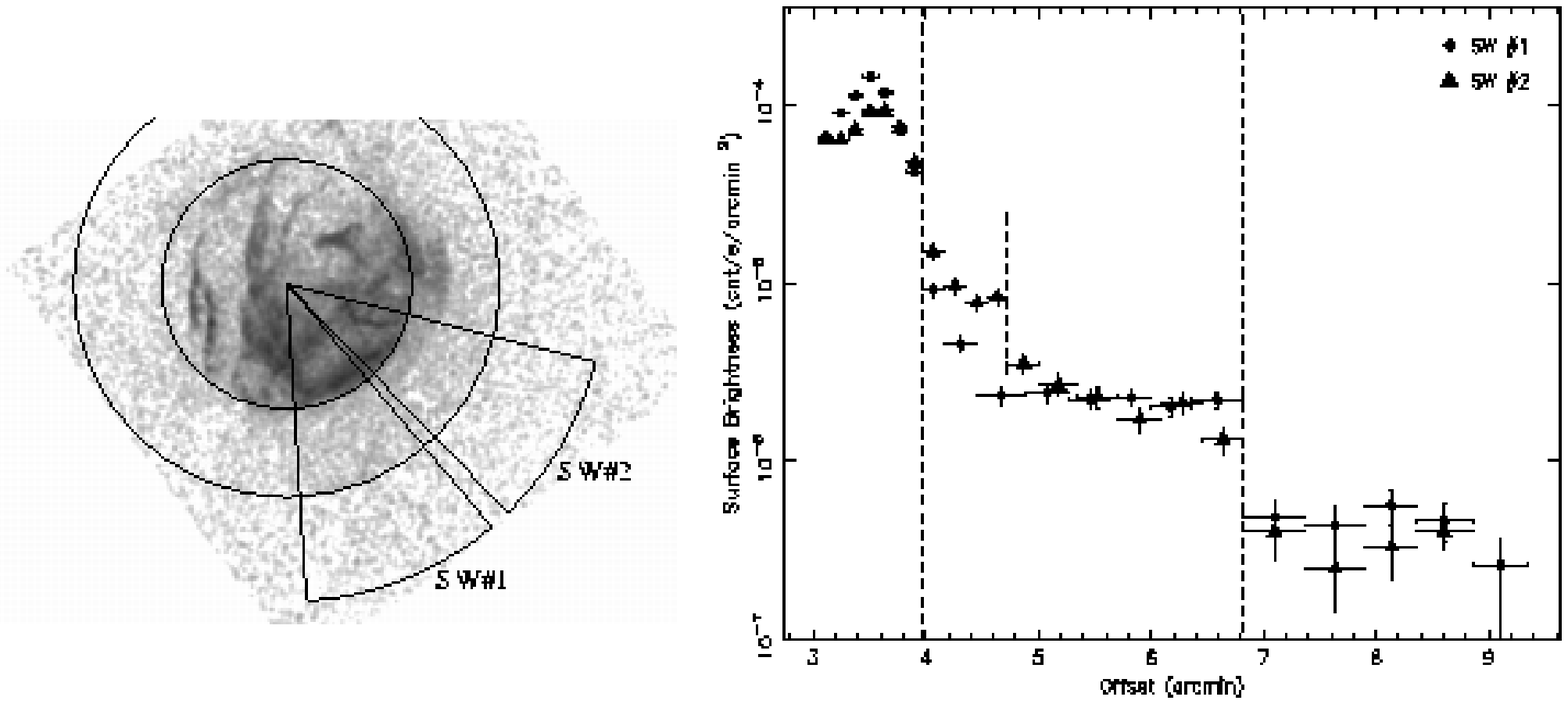}}
\vspace{-5cm}
  \caption{The radial surface brightness profiles of two SW sectors. Both
sectors have two common surface brightness discontinuities (the inner
shell and the outer shell), while SW\#2 has an additional
discontinuity, the middle shell indicated on the right.
    \label{fig:img:smo}}
\end{figure*}
%%%%%%%%%%%%%%%%%%%%%%%%%%%%%%%%%%%%%%%%%%%%%%%%%%%%%%%%%%%%%%%%%%%%%%%%%%
\clearpage

%%%%%%%%%%%%%%%%%%%%%%%%%%%%%%%%%%%%%%%%%%%%%%%%%%%%%%%%%%%%%%%%%%%%%%%%%%
\begin{inlinefigure}
\vspace{1.5cm}
  \centerline{\includegraphics[height=0.7\linewidth,angle=90]{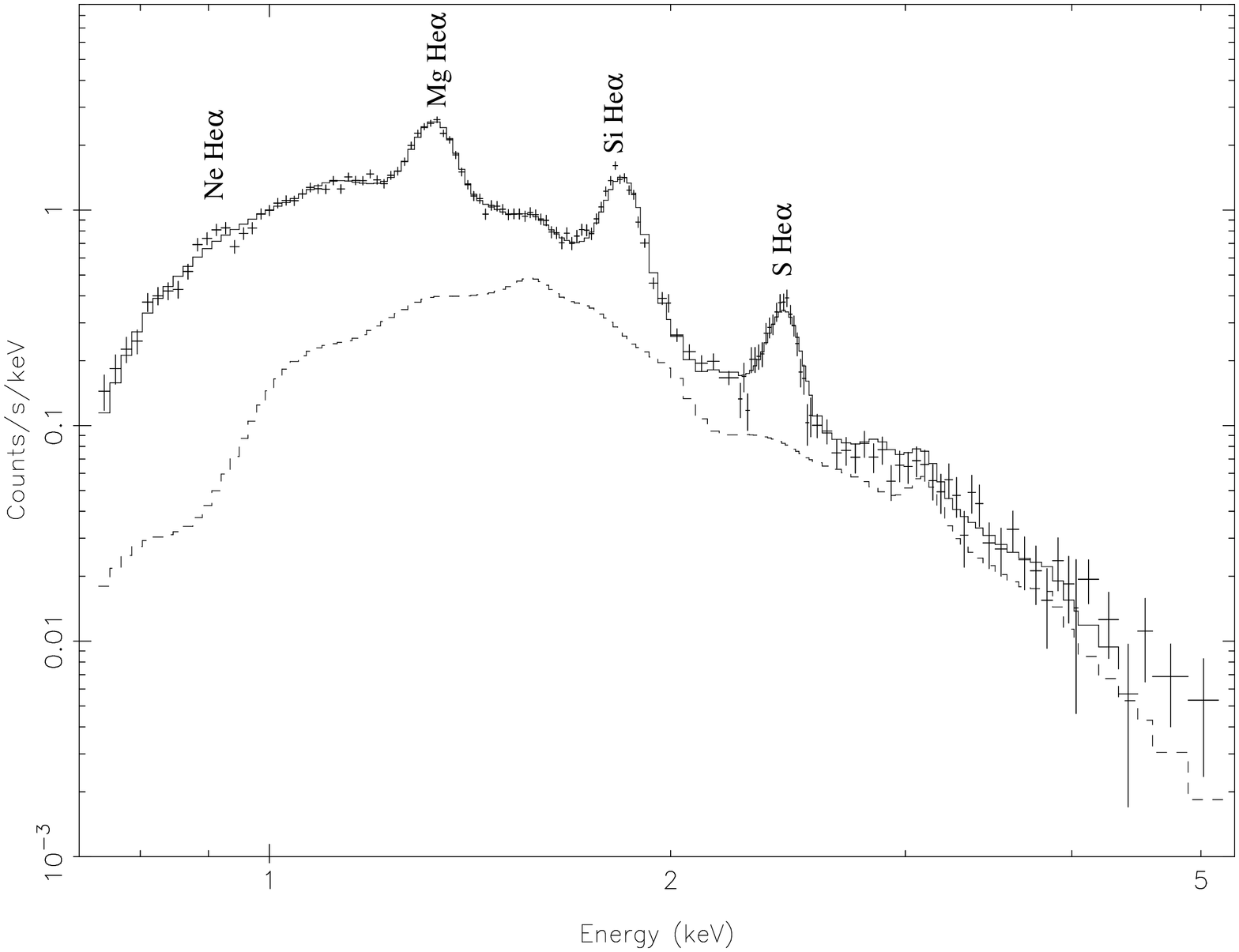}}
\vspace{-0.5cm}
  \caption{The integrated \chandra\ spectrum of the whole remnant with the
best-fit NEI model (SPEX 2.0). Emission lines from Ne, Mg, Si and S are
labeled. Fe L-blend is also visible between Ne and Mg
lines. The dashed line represents the residual emission without the
contribution from Ne, Mg, Si, S and Fe.
    \label{fig:img:smo}}
\end{inlinefigure}
%%%%%%%%%%%%%%%%%%%%%%%%%%%%%%%%%%%%%%%%%%%%%%%%%%%%%%%%%%%%%%%%%%%%%%%%%%

%%%%%%%%%%%%%%%%%%%%%%%%%%%%%%%%%%%%%%%%%%%%%%%%%%%%%%%%%%%%%%%%%%%%%%%%%%
\begin{figure*}[htb]
\vspace{-7cm}
  \centerline{\includegraphics[height=1.4\linewidth]{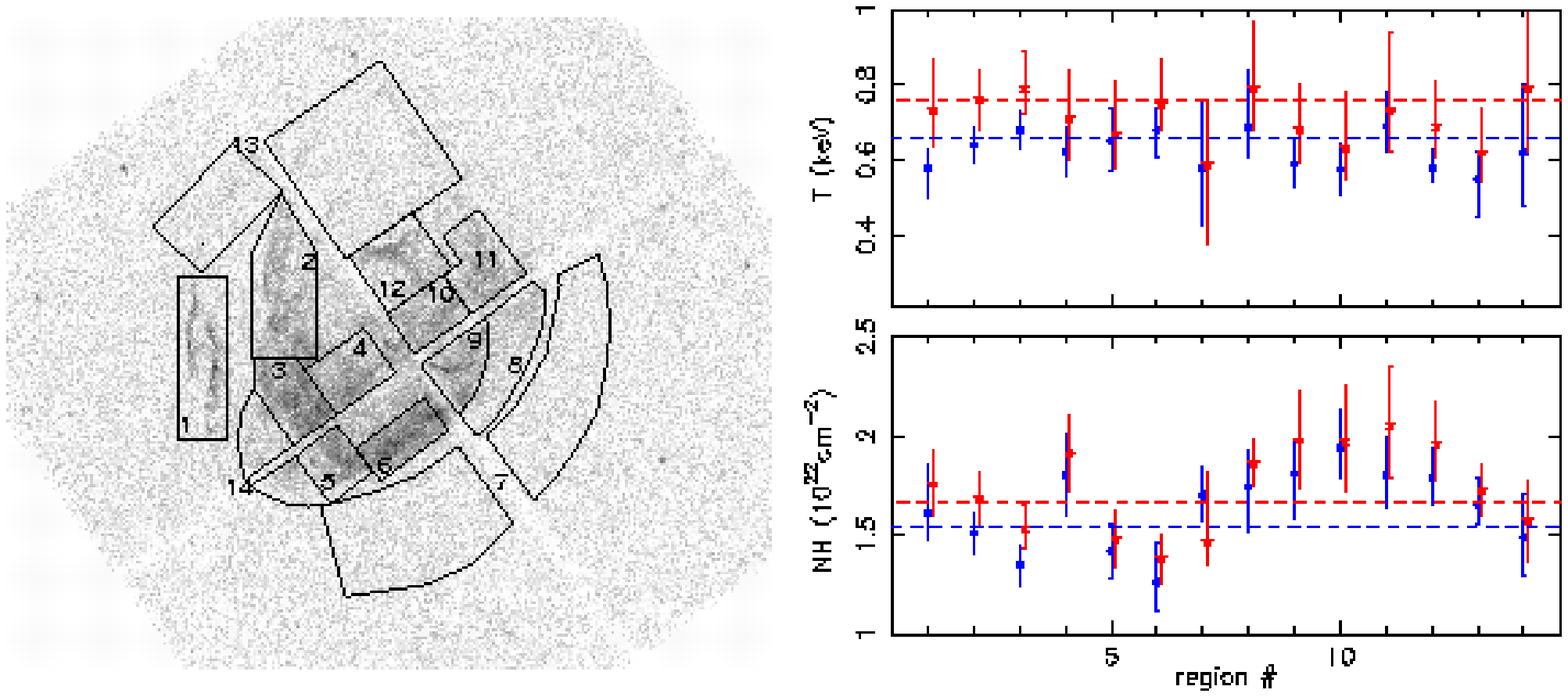}}
\vspace{-6cm}
  \caption{The measured temperatures and absorption in 14 regions of Kes
79. The blue points are the results of XSPEC VNEI fits, while the red points
are the results of SPEX NEI fits. The dashed lines represent the best fits
to the \chandra\ global spectrum. Regardless of the different results
from two NEI codes, the temperature variation across the remnant is small,
which implies a nearly isothermal state of plasma in Kes 79. The Northwest
part of the remnant has higher absorption than the Southeast part.
    \label{fig:img:smo}}
\end{figure*}
%%%%%%%%%%%%%%%%%%%%%%%%%%%%%%%%%%%%%%%%%%%%%%%%%%%%%%%%%%%%%%%%%%%%%%%%%%
\clearpage

\end{document}